\documentclass[twocolumn,showpacs,aps,prl,superscriptaddress,preprintnumbers,floatfix]{revtex4}

\usepackage{graphicx}
\usepackage{dcolumn}
\usepackage{amsmath}
\usepackage{epsfig}
\usepackage{relsize}

\RequirePackage{xspace}
\def\invfb    {\ensuremath{\mbox{\,fb}^{-1}}\xspace}
\def\pep2{PEP-II}
\def\epem     {\ensuremath{e^+e^-}\xspace}
\def\ccbar    {\ensuremath{c\overline c}\xspace}
\def\CP       {\ensuremath{C\!P}\xspace}
\def\CPT      {\ensuremath{C\!PT}\xspace}
\def\T        {\ensuremath{T}\xspace}
\def\to       {\ensuremath{\rightarrow}\xspace}
\mathchardef\Upsilon="7107
\def\Y#1S{\ensuremath{\Upsilon{(#1S)}}\xspace}

\def\TwoS  {\Y2S}
\def\ThreeS{\Y3S}
\def\FourS {\Y4S}

\def\Dstarp   {\ensuremath{D^{*+}}\xspace}
\def\Dp       {\ensuremath{D^+}\xspace}
\def\Dm       {\ensuremath{D^-}\xspace}
\def\Ds       {\ensuremath{D_s^+}\xspace} 
\def\Dbar     {\kern 0.2em\overline{\kern -0.2em D}{}\xspace}
\def\Dz       {\ensuremath{D^0}\xspace}
\def\Dzb      {\ensuremath{\Dbar^0}\xspace}
\def\Km       {\ensuremath{K^-}\xspace}
\def\Kp       {\ensuremath{K^+}\xspace}
\def\pim      {\ensuremath{\pi^-}\xspace}
\def\pip      {\ensuremath{\pi^+}\xspace}
\def\KS       {\ensuremath{K^0_S}\xspace}
\def\proton   {\ensuremath{p}\xspace}
\def\todd     {\ensuremath{T}-odd\xspace}
\def\DpDecay  {\ensuremath{\Dp \to \Kp \KS \pip \pim}\xspace}
\def\DsDecay  {\ensuremath{\Ds \to \Kp \KS \pip \pim}\xspace}
\def\DpRef    {\ensuremath{\Dp \to \KS \pip \pip \pim}\xspace}
\def\LamRef   {\ensuremath{\Lambda_c^+ \to \proton \KS \pip \pim}\xspace}
\def\DsBest   {\ensuremath{\Ds \to \KS \Km \pip \pip}\xspace}
\def\Z        {\ensuremath{Z^0}\xspace}
\def\Dps      {\ensuremath{D^+_{(s)}}\xspace}
\def\Dms      {\ensuremath{D^-_{(s)}}\xspace}

\def\Dsm      {\ensuremath{D^-_{s}}\xspace}
\def\Ct       {\ensuremath{C_T}\xspace}
\def\Ctb      {\ensuremath{\bar{C}_T}\xspace}
\def\At       {\ensuremath{A_T}\xspace}
\def\Atbar    {\ensuremath{\bar{A}_T}\xspace}
\def\Atv      {\ensuremath{\mathcal{A}_T}\xspace}
\def\dm       {\ensuremath{\Delta m}\xspace}
\def\gstar    {\ensuremath{\gamma^*}\xspace}
\def\jetset   {\mbox{\tt Jetset \hspace{-0.5em}7.\hspace{-0.2em}4}\xspace}
\def\geantFour{\mbox{\tt GEANT \hspace{-0.5em}4}\xspace}
\def\stat     {\mbox{stat}\xspace}
\def\syst     {\mbox{syst}\xspace}
\def\babar{\mbox{\sl B\hspace{-0.4em} {\small\sl A}\hspace{-0.37em} \sl B\hspace{-0.4em} {\small\sl A\hspace{-0.02em}R}}}  

\newcommand{\gev}{\ensuremath{\mathrm{\,Ge\kern -0.1em V}}\xspace}
\newcommand{\gevc}{\ensuremath{{\mathrm{\,Ge\kern -0.1em V\!/}c}}\xspace}
\newcommand{\gevcc}{\ensuremath{{\mathrm{\,Ge\kern -0.1em V\!/}c^2}}\xspace}
\newcommand{\mevcc}{\ensuremath{{\mathrm{\,Me\kern -0.1em V\!/}c^2}}\xspace}

\newcommand{\BABARPubYear}    {11}
\newcommand{\BABARPubNumber}  {007}

\newcommand{\SLACPubNumber} {14464}
\newcommand{\LANLNumber} {1105.4410}

\def\figurebox#1#2#3{%
    \def\arg{#3}%
    \ifx\arg\empty
    {\hfill\vbox{\hsize#2\hrule\hbox to #2{\vrule\hfill\vbox to #1{\hsize#2\vfill}\vrule}\hrule}\hfill}%
    \else
    {\hfill\epsfbox{#3}\hfill}%
    \fi}

\begin{document}

\preprint{\babar\  Analysis Document \#2394, Version 15}

\begin{flushleft}
\babar-PUB-\BABARPubYear/\BABARPubNumber\\
SLAC-PUB-\SLACPubNumber\\
arXiv:\LANLNumber\ [hep-ex]\\[10mm]
\end{flushleft}

\title{
{\large \bf
Search for \CP violation using $T$-odd correlations in \DpDecay and \DsDecay decays.} 
}

%
\author{J.~P.~Lees}
\author{V.~Poireau}
\author{V.~Tisserand}
\affiliation{Laboratoire d'Annecy-le-Vieux de Physique des Particules (LAPP), Universit\'e de Savoie, CNRS/IN2P3,  F-74941 Annecy-Le-Vieux, France}
\author{J.~Garra~Tico}
\author{E.~Grauges}
\affiliation{Universitat de Barcelona, Facultat de Fisica, Departament ECM, E-08028 Barcelona, Spain }
\author{M.~Martinelli$^{ab}$}
\author{D.~A.~Milanes$^{a}$}
\author{A.~Palano$^{ab}$ }
\author{M.~Pappagallo$^{ab}$ }
\affiliation{INFN Sezione di Bari$^{a}$; Dipartimento di Fisica, Universit\`a di Bari$^{b}$, I-70126 Bari, Italy }
\author{G.~Eigen}
\author{B.~Stugu}
\author{L.~Sun}
\affiliation{University of Bergen, Institute of Physics, N-5007 Bergen, Norway }
% ...
\collaboration{The \babar\ Collaboration}
\noaffiliation

\date{\today}

\begin{abstract}

We search for \CP violation in a sample of 20000 Cabibbo-suppressed decays, \DpDecay, and 30000 Cabibbo-favored decays, \DsDecay. 
We use 520\invfb of data recorded by the \babar\ detector at the \pep2 asymmetric-energy \epem collider operating at center of mass energies near 10.6 \gev.
We search for \CP violation in the difference between the $T$-odd asymmetries obtained using triple product correlations of the \Dp (\Ds) and \Dm (\Dsm) decays, respectively.
The \T violation parameter values obtained are $\Atv(\Dp) = (-12.0 \pm 10.0_{\stat} \pm 4.6_{\syst})\times 10^{-3}$ and $\Atv(\Ds) = (-13.6 \pm 7.7_{\stat} \pm 3.4_{\syst}) \times10^{-3}$, 
which are consistent with the standard model expectations. 
\end{abstract}

\maketitle
In the standard model (SM) of particle physics, the violation of the charge-conjugation and parity symmetries (\CP) is introduced by the Kobayashi-Maskawa (KM) phase in the Cabibbo-Kobayashi-Maskawa quark mixing matrix~\cite{CKM}. 
The KM ansatz has been tested at high precision in $K$ and $B$ decays, where the KM phase contributes to the quark transition amplitude at tree level.
However, further experimental efforts are needed in $D$ meson decays, where \CP-violating amplitudes are predicted to contribute to the observables at the $10^{-3}$ level~\cite{Buccella}.

The sensitivity to \CP violation in $D$ meson decays reached by the $B$ factories is of the order of $5\times10^{-3}$~\cite{delAmoSanchez:2010xj,delAmoSanchez:2011xx,Ko:2011xx,Ko:2010ng}.
Although this does not represent a measurement of SM \CP violation, it provides a constraint on possible effects beyond the SM.
New physics models introduce \CP violation in $D$ meson decays through tree and one-loop diagrams. 
While predictions for \CP violation in tree diagrams are not different from those in the SM [$\mathcal{O}(10^{-3}$)], new physics in loop diagrams may enhance \CP violation effects at the order of $10^{-2}$~\cite{Grossman:2006jg}.

We report herein a search for \CP violation in the decays \DpDecay and \DsDecay using \todd correlations~\cite{conj}.
We define a kinematic triple product that is odd under time reversal using the vector momenta of the final state particles in the \Dps rest frame as
\begin{equation}
\Ct \equiv \vec{p}_{\Kp} \cdot \left( \vec{p}_{\pip} \times \vec{p}_{\pim} \right).
\label{eq:Ct}
\end{equation}
Under the assumption of \CPT invariance, time-reversal (\T) violation is equivalent to \CP violation.

We study the \todd correlations by measuring the observable expressed in Eq.~(\ref{eq:Ct}) and then evaluating the asymmetry
\begin{equation}
\At \equiv \frac{\Gamma(\Ct>0) - \Gamma(\Ct<0)}{\Gamma(\Ct>0) + \Gamma(\Ct<0)},
\label{eq:At}
\end{equation}
where $\Gamma$ is the decay rate for the process under study.
The observable defined in Eq.~(\ref{eq:At}) can have a nonzero value due to final state interactions, even if the weak phases are zero~\cite{Bigi:2009zzb}.
The \todd asymmetry measured in the \CP-conjugate decay process, \Atbar, is defined as
\begin{equation}
\Atbar \equiv \frac{\Gamma(-\Ctb>0) - \Gamma(-\Ctb<0)}{\Gamma(-\Ctb>0) + \Gamma(-\Ctb<0)},
\label{eq:Atb}
\end{equation}
where $\Ctb\equiv \vec{p}_{\Km} \cdot \left( \vec{p}_{\pim} \times \vec{p}_{\pip} \right)$.
We can then construct
\begin{equation}
\Atv \equiv \frac{1}{2}\left( \At - \Atbar \right),
\label{eq:Atv}
\end{equation}
which is an asymmetry that characterizes \T violation in the weak decay process~\cite{Bensalem:2002ys,Bensalem:2002pz,Bensalem:2000hq}.

At least four different particles are required in the final state so that the triple product may be defined using momentum vectors only~\cite{Golowich:1988ig}.
The $D$ meson decays suitable for this analysis method are \DpDecay, \DsDecay, and $\Dz\to\Kp\Km\pip\pim$. 
The search for \CP violation using \todd correlations in $\Dz\to\Kp\Km\pip\pim$ has recently been carried out by the \babar\  Collaboration, and no evidence of \CP violation has been observed~\cite{delAmoSanchez:2010xj}.

Following the suggestion by Bigi~\cite{Bigi:2001sg}, the FOCUS Collaboration~\cite{Link:2005th} first applied this technique to a sample of approximately 500 reconstructed \Dp and \Ds events, respectively. 
No evidence of \CP violation was found.
In the present analysis, we perform a similar measurement using approximately $2.1\times10^4$ \Dp and $3.0\times10^4$ \Ds meson decay candidates.

The analysis is based on a 520~$\invfb$ data sample recorded mostly at the \FourS peak and at center of mass (CM) energy 40 MeV below the resonance by the \babar\ detector at the \pep2 asymmetric-energy \epem collider.  
Contributions to the data sample have been recorded near the \ThreeS resonance ($\approx31\invfb$), and near the \TwoS resonance ($\approx15\invfb$).
In addition, two large samples of Monte Carlo (MC) simulated events have been analyzed.
In these samples, the $\epem\to\ccbar$ production process is generated using \jetset~\cite{jetset}, and the detector response is simulated by \geantFour~\cite{geant}.
About $1.1\times10^9$ generic $\epem\to\ccbar$ MC events, corresponding to 846~\invfb, were generated to include the previously measured intermediate resonances in the \Dps decays, while $4.0\times10^6$ $\epem \to \Dps X$ MC signal events ($\approx 1025\invfb$), where $X$ represents any system of charged and neutral particles compatible with the relevant conservation laws, were generated in which the \Dps signal decays to $\Kp\KS\pip\pim$ uniformly over the phase space. 
Both MC samples were processed using the same reconstruction and analysis chain as that used for real events.

The \babar\ detector is described in detail elsewhere~\cite{babar}. 
We mention here only the subsystems used in the present analysis.
Charged-particle tracks are detected, and their momenta measured, with a combination of a cylindrical drift chamber (DCH) and a silicon vertex tracker (SVT), both operating within the 1.5-T magnetic field of a superconducting solenoid.
The information from a ring-imaging Cherenkov detector, combined with specific energy-loss measurements in the SVT and DCH, provides identification of charged kaon and pion candidates. 

The \Dp and \Ds meson decay candidates are reconstructed in the production and decay sequence:
\begin{equation}
\epem\to X\Dps; \Dps\to\Kp\KS\pip\pim; \KS\to\pip\pim,
\label{eq:reaction}
\end{equation}
using the events with at least five charged particles.
We reconstruct $\KS\to\pip\pim$ candidates using a vertex and kinematic fit with the \KS mass constraint~\cite{Nakamura:2010zzi}, and requiring a $\chi^2$ probability greater than 0.1\%.  
We accept only \KS candidates that decay at least 0.5 cm from the \epem interaction region (IR) and have a mass before the fit within 15\mevcc of the nominal \KS mass.
The \KS candidate is then combined with three charged-particle tracks with total net charge $+1$, to form a \Dps candidate.
We require the tracks to originate from a common vertex, and the $\chi^2$ fit probability ($P_1$) to be greater than 0.1\%. 
In order to improve discrimination between signal and background, an additional fit is performed that constrains the three charged tracks to the IR. 
The $\chi^2$ probability ($P_2$) of this fit is large for most of the background events, whose tracks originate from the IR, while it is smaller for \Dps signal events, whose tracks originate from a secondary vertex detached from the IR, due to the measurable \Dps flight distance.
Particle identification is applied to the three charged-particle tracks, and the presence of a \Kp is required.
Charged kaon identification has an average efficiency of 90\% within the acceptance of the detector, and an average pion-to-kaon misidentification probability of 1.5\%. 
We require the CM momentum of the \Dps candidate, $p^*$, to be greater than 2.5\gevc. 
This requirement reduces the large combinatorial background from $B$ decays, and improves the signal-to-background ratio significantly despite some loss in signal efficiency.

We first study backgrounds from charm meson decay processes which yield the same event topology.

The decay $\Dstarp\to\pip\Dz$ produces a significant \Dz peak in the $\KS\Kp\pim$ mass distribution.
A fit with a Gaussian signal function yields a mass resolution of $\sigma_{\Dz \to \KS\Kp\pim}=4.6$~\mevcc.
Selecting \Dz candidates within $\pm3\sigma_{\Dz \to \KS\Kp\pim}$ of the \Dz mass, we observe a clear \Dstarp peak in the  distribution of the mass difference $\dm=m(\Kp\KS\pip\pim) - m(\KS\Kp\pim)$. 
This contribution is reduced to a negligible level by requiring $\dm> 0.1465$~\gevcc.

We also observe background from the decay $\Dp\to\Kp\KS\KS$, with one of the \KS decaying into the bachelor pions of Eq.~(\ref{eq:reaction}).
This contribution is removed by requiring the $\pip\pim$ invariant mass to lie outside a $\pm 8.7$ \mevcc mass window around the nominal \KS mass~\cite{Nakamura:2010zzi}.
We look for backgrounds from \DpRef decays by assigning a pion mass hypothesis to the kaon candidate.
We observe a \Dp signal over a large background. 
Simulation shows that this background produces a broad structure in the high-mass region of the \Ds mass distribution.
We also looked for background from \LamRef decay by assigning the proton mass to the kaon candidate.
We see a signal over a large background.
We find it impossible to remove the \DpRef and \LamRef events without biasing our mass distributions. 
Our MC simulations, however, show that the presence of these backgrounds does not bias the extraction of the \Dps meson yields.
As a further check, we select a high purity data sample (87.5 \%) of \DpRef decays and assign the \Kp mass alternatively to both \pip. 
We compute the asymmetries on the resulting integrated distributions and find that they are all consistent with zero. 
A similar result is obtained when we perform the test on MC events.
 
We divide the $\Kp\KS\pip\pim$ mass spectrum into two regions in order to extract separately the \Dp and \Ds signal yields. 
For the former we require $1.81 < m(\Kp\KS\pip\pim) < 1.92\gevcc$, while for the latter we require $1.91 < m(\Kp\KS\pip\pim) < 2.02\gevcc$.

For further signal-to-background optimization, we explore three variables: the CM momentum, $p^*$, 
the difference in probability, $P_1 - P_2$, and the signed transverse decay length, $L_T = \frac{\vec d \cdot \vec p_T}{|\vec p_T|}$, where $\vec d$ is the distance vector between the IR and the \Dps decay vertex in the transverse plane, and $\vec p_T$ is the \Dps transverse momentum vector.
Signal events are expected to be characterized by larger values of $p^*$~\cite{Aubert:2002ue}, due to the jetlike topology of $\epem\to\ccbar$ events, and larger values of $L_T$ and $P_1-P_2$, due to the measurable \Dps decay length.

\begin{figure*}
\includegraphics[width=0.8\textwidth]{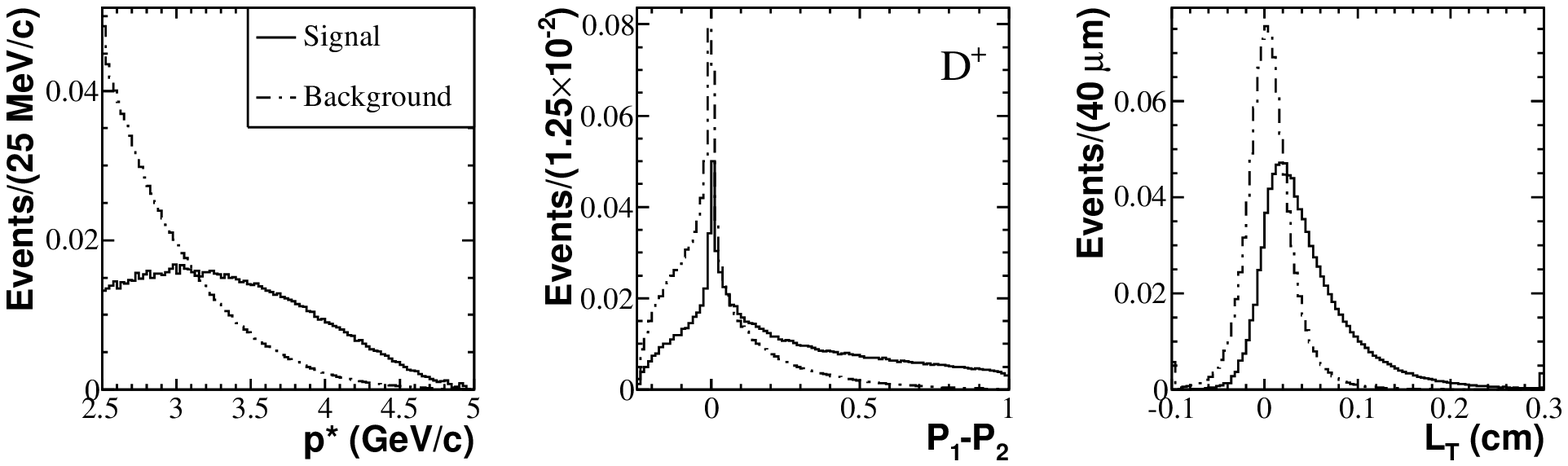}
\includegraphics[width=0.8\textwidth]{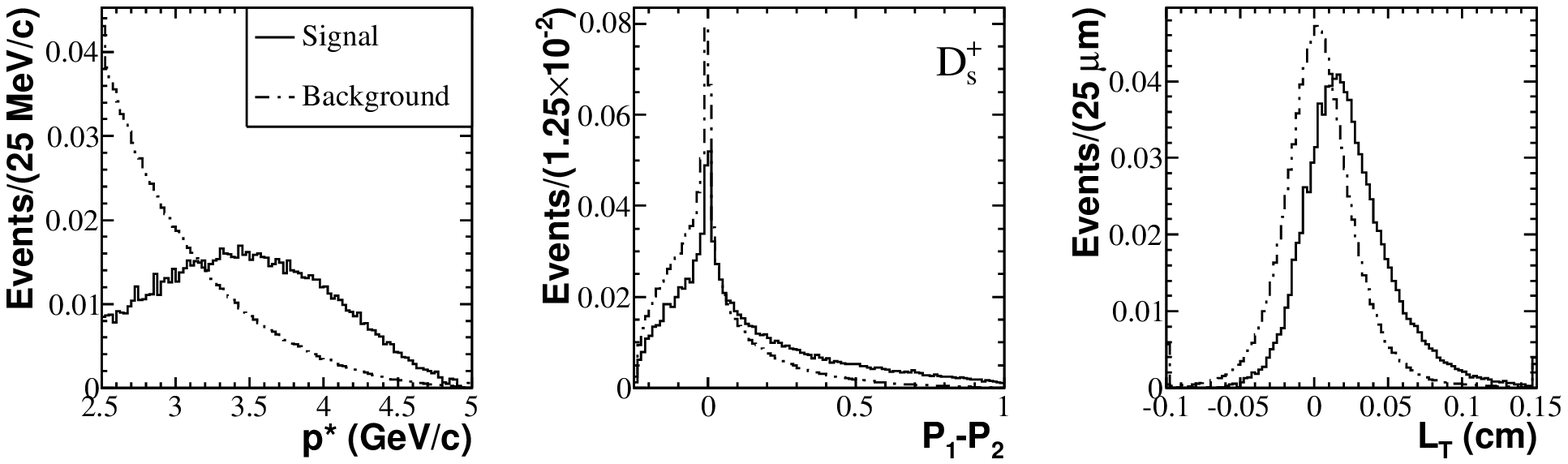}
\caption{\label{fig:LRdist} Distributions of $p^*$, $P_1-P_2$, and $L_T$ for \Dp (top panels) and \Ds (bottom panels) candidates. 
The distributions for signal and background are shown as solid and dot-dashed histograms, respectively.
All distributions are normalized to 1.
Signal distributions are extracted from \DpRef and \DsBest for \Dp and \Ds decays, respectively.
The background distributions are extracted from the $\Dps\to\Kp\KS\pip\pim$ sidebands.}
\end{figure*}

Figure~\ref{fig:LRdist} shows the $p^*$, $P_1-P_2$, and $L_T$ distributions for signal and background in the \Dp and \Ds mass regions, respectively.
The signal distributions are obtained from \DpRef and \DsBest decays in data after background subtraction.
These decay modes are kinematically similar to the signal modes, but have higher signal yields and better signal-to-background ratios.
The background distributions in Fig.~\ref{fig:LRdist} are obtained from $\Dps\to\Kp\KS\pip\pim$ sidebands in the mass distributions for data.

The normalized probability distribution functions ($\mathcal{P}$) of the three variables for signal and background are combined in a likelihood-ratio test 
\begin{align}
\mathcal{L}=\prod_i\frac{\mathcal{P}^{sig}_i(x_i)}{\mathcal{P}^{bkg}_i(x_i)},\qquad\vec{x}=(p^*,P_1-P_2,L_T)
\end{align} 
to optimize the signal yields separately for \Dp and \Ds.
The optimization of the cut is performed by maximizing the value of $S/\sqrt{S+B}$, where $S$ is the number of signal events and $S+B$ is the total number of events in the signal region.
The purity $S/(S+B)$ of the peak improves from 11.2\% to 51.4\% and from 16.6\% to 60.6\% for \Dp and \Ds, respectively.

\begin{figure}
\includegraphics[width=0.23\textwidth]{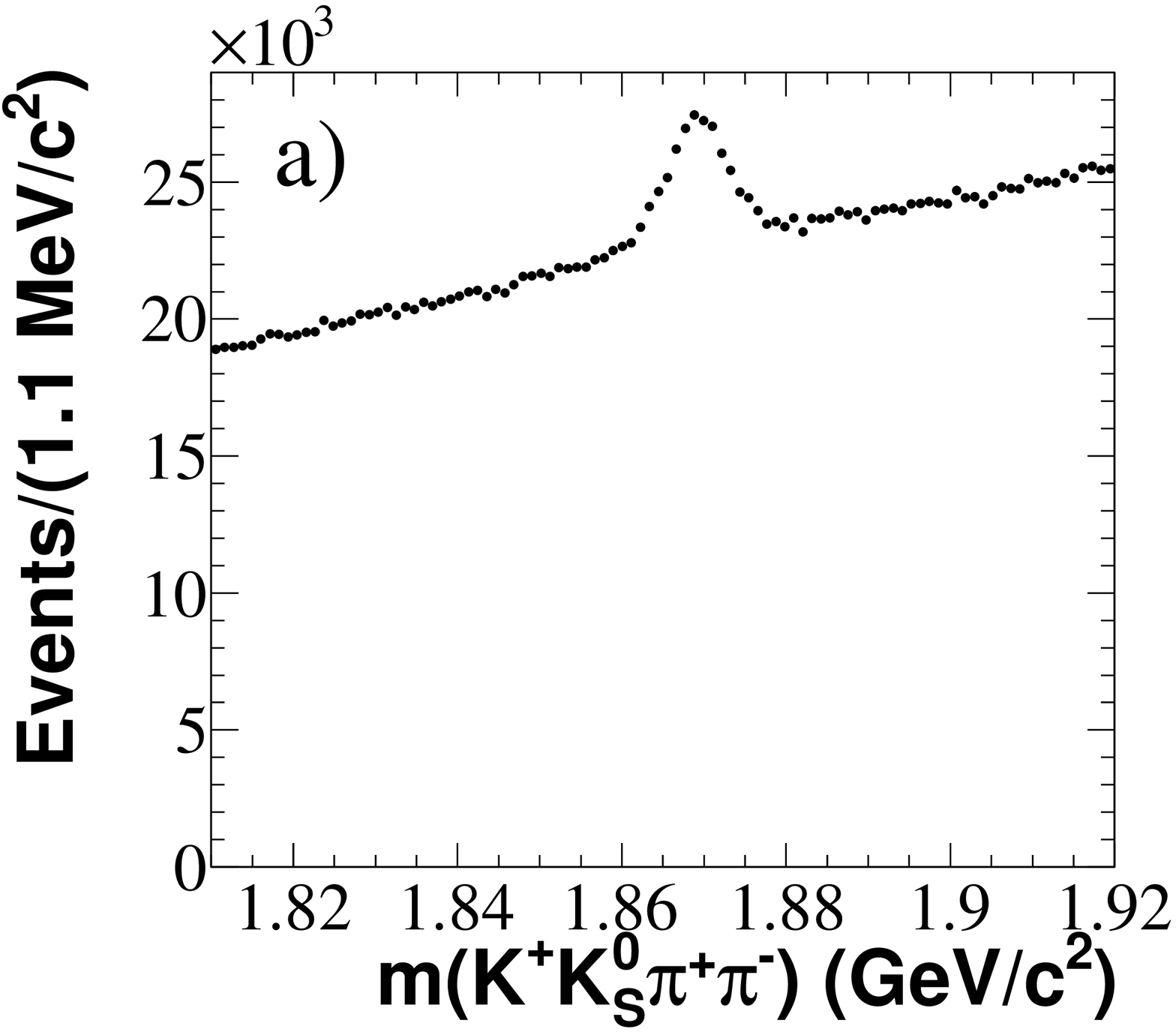}
\includegraphics[width=0.23\textwidth]{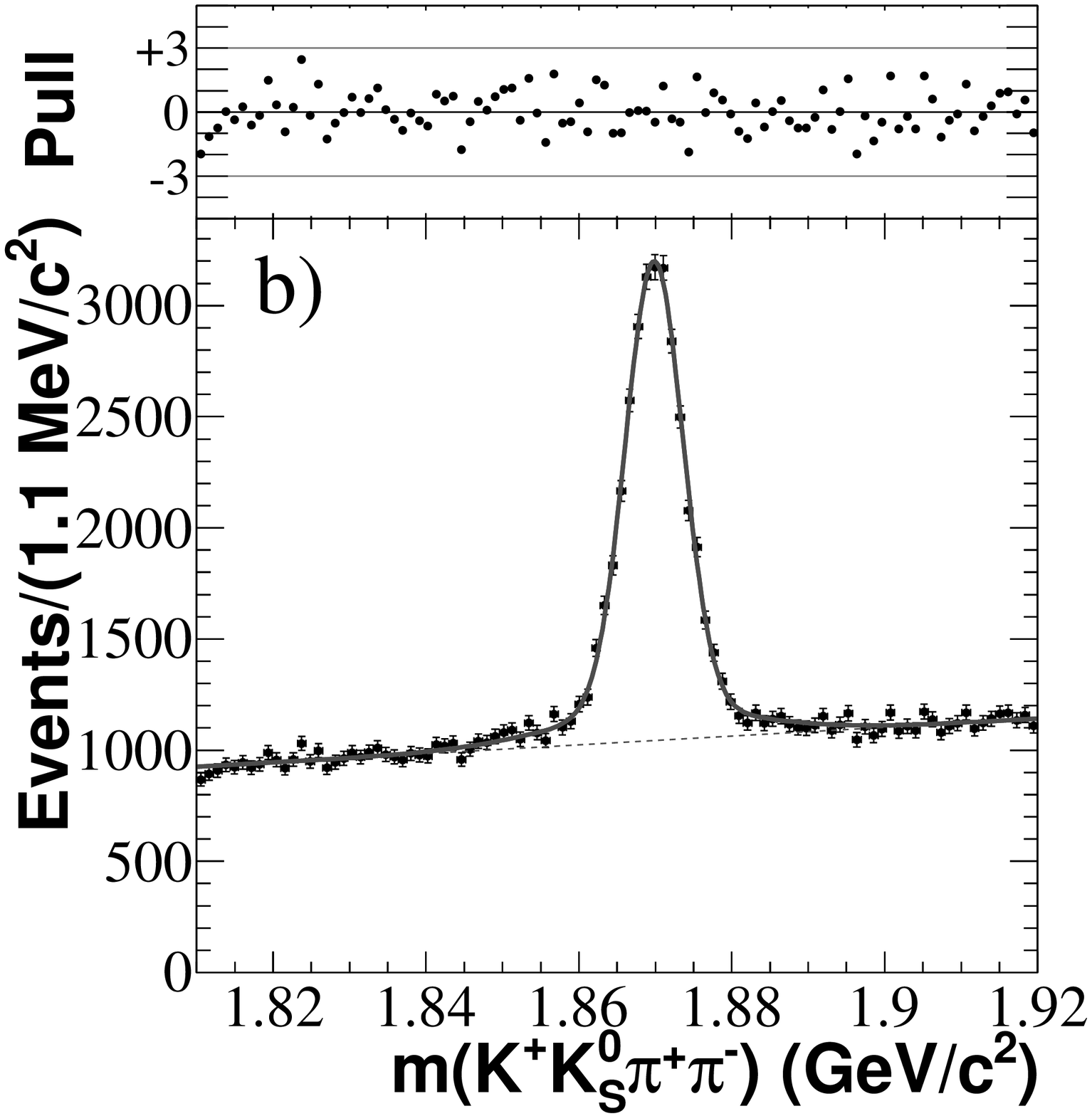}
\includegraphics[width=0.23\textwidth]{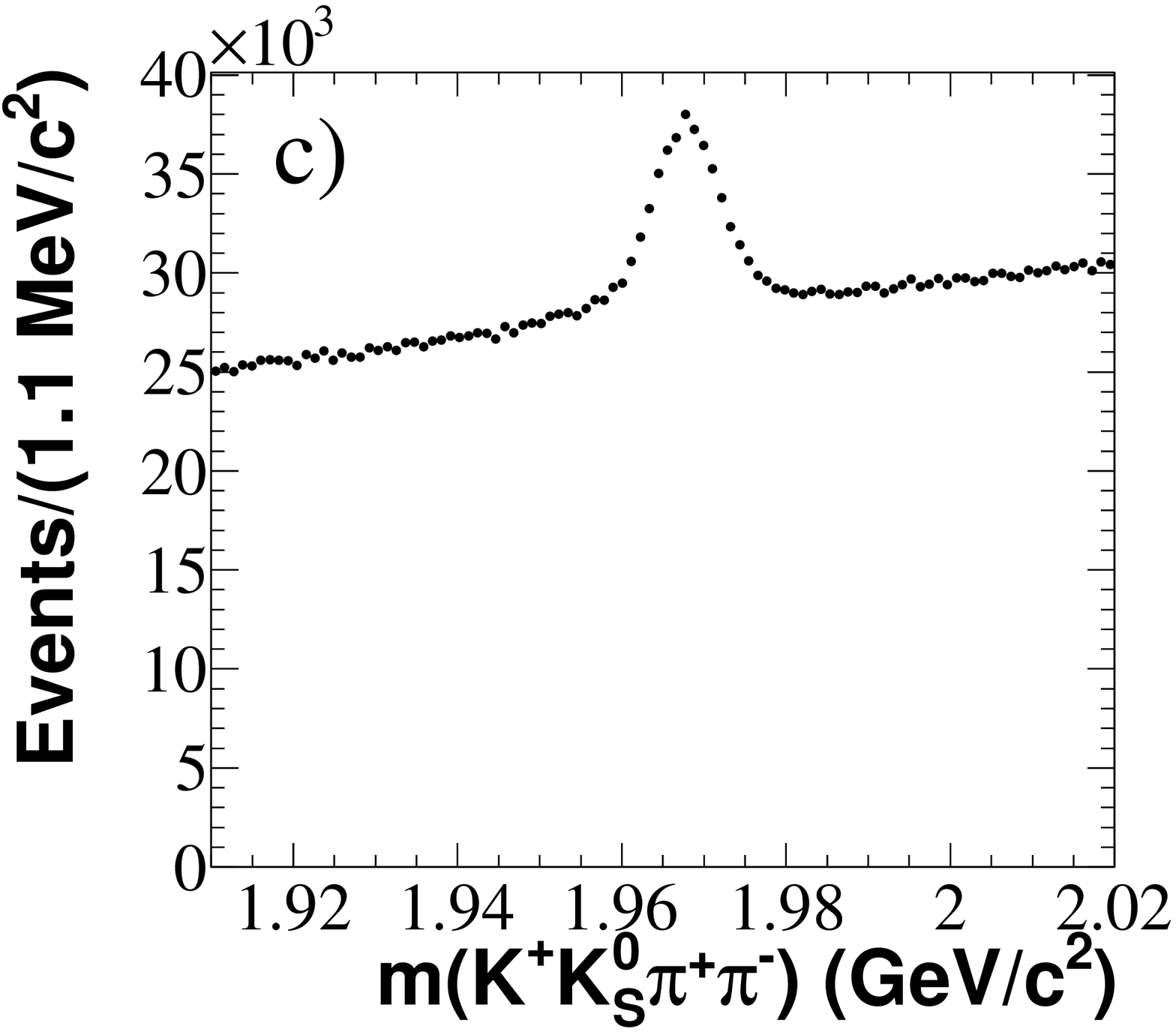}
\includegraphics[width=0.23\textwidth]{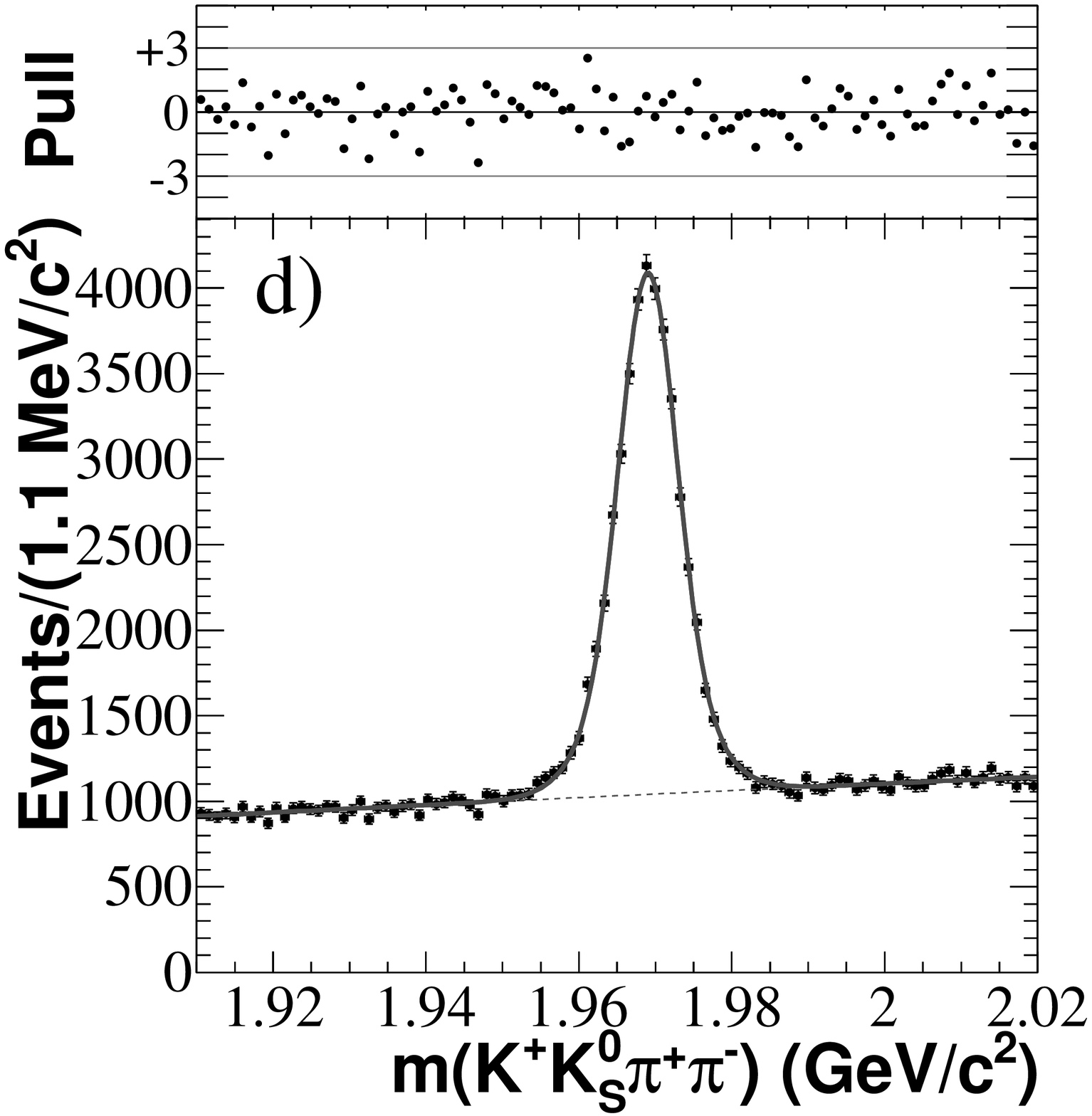}
\caption{\label{fig:fig2} The $\Kp\KS\pip\pim$ mass spectrum in the \Dp mass region (a) before and (b) after the cut on likelihood ratio.
Similar plots (c) and (d) are drawn for \Ds.
The curves in (b) and (d) result from the fits described in the text. 
The distributions of the pull values are also shown.
The $\chi^2/n_{dof}$ values from the fits are 0.87 (\Dp) and 0.95 (\Ds).}
\end{figure}

Figure~\ref{fig:fig2} shows the $\Kp\KS\pip\pim$ mass spectra in the \Dp and \Ds regions before [(a) and (c)] and after [(b) and (d)] the likelihood-ratio test.
For each region, the signal is described by the superposition of two Gaussian functions with a common mean value.
The background is parametrized by a first-order polynomial in the \Dp region, and by a second-order polynomial in the \Ds region. 
The fitted functions are superimposed on the data in Fig.~\ref{fig:fig2}, and the fit residuals, shown above each distribution, are represented by $\text{Pull} = (N_{\text{data}} - N_{\text{fit}})/\sqrt{N_{\text{data}}}$.
From these binned extended maximum likelihood fit, we extract the integrated yields $N(\Dp) = 21210 \pm 392$ and $N(\Ds) = 29791 \pm 337$ from the fits, where the uncertainties are statistical only.
The mean value and width of the main Gaussian are $\mu_{\Dp}=1869.8\pm0.1\mevcc$, $\sigma_{\Dp}=3.76\pm0.08\mevcc$ for \Dp, and $\mu_{\Ds}=1969.0\pm0.1\mevcc$, $\sigma_{\Ds}=3.67\pm0.18\mevcc$ for \Ds.

We next divide the data sample into four subsamples depending on $D_{(s)}$ charge and whether \Ct (\Ctb) is greater or less than zero.
We define
\begin{align}
\nonumber 	N(\Dps, \Ct>0) 	&= \frac{N(\Dps)}{2}\left( 1 + \At \right),\\
\nonumber 	N(\Dps, \Ct<0) 	&= \frac{N(\Dps)}{2}\left( 1 - \At \right),\\
\nonumber 	N(\Dms,\Ctb>0)	&= \frac{N(\Dms)}{2}\left( 1 - \Atbar \right),\\
			N(\Dms,\Ctb<0)	&= \frac{N(\Dms)}{2}\left( 1 + \Atbar \right),
\end{align}
and fit the corresponding mass spectra simultaneously to extract the yields and the values of the asymmetry parameters \At and \Atbar. 
In this fit, the shape parameters are shared among the four samples and are fitted together with the yields, $N(\Dps)$ and $N(\Dms)$, and the asymmetries, \At and \Atbar.
The \T-violating parameter \Atv is then computed using Eq.~(\ref{eq:Atv}).

We validate the method by using the generic MC sample. 
We find that the fit results for \At, \Atbar, and the computed value of \Atv are in good agreement with those in the simulation, both for \Dp and \Ds.

All event selection criteria are determined before the final fit in order to avoid any potential bias.
The true central values of \At and \Atbar are masked by adding unknown random offsets. 

After removing the offsets, we measure the following asymmetries:
\begin{align}
\nonumber 	\At (\Dp)		&= (+11.2\pm14.1_{\stat}\pm 5.7_{\syst})\times10^{-3},\\
			\Atbar(\Dm)	&= (+35.1\pm14.3_{\stat}\pm 7.2_{\syst})\times10^{-3},
\label{eq:AtRes}
\end{align}
and
\begin{align}
\nonumber 	\At(\Ds)		&= (-99.2\pm10.7_{\stat}\pm  8.3_{\syst})\times10^{-3},\\
			\Atbar(\Dsm) 	&= (-72.1\pm10.9_{\stat}\pm 10.7_{\syst})\times10^{-3}.
\label{eq:AtbRes}
\end{align}

We observe values of $A_T$ and $\overline{A}_T$ which differ significantly from zero only for \Ds decay. 
This may indicate the presence of  final-state-interaction effects for this decay process, perhaps as a result of the slightly different resonant substructure between \Dp and \Ds decay.
For example, the $K^{*0}K^{*+}$ final state can contribute only to \Ds through a doubly Cabibbo-suppressed decay process.
In the case of \Dp decay we find  $A_T$ and $\overline{A}_T$ to be consistent with zero, in contrast with the results of a similar analysis performed on the corresponding \Dz decay sample~\cite{delAmoSanchez:2010xj}:
\begin{align}
\nonumber 	\At(\Dz)		&= (-68.5\pm 7.3_{\stat}\pm 5.8_{\syst})\times10^{-3},\\
			\Atbar(\Dzb) 	&= (-70.5\pm 7.3_{\stat}\pm 3.9_{\syst})\times10^{-3}.
\label{eq:AtDz}
\end{align}	

\begin{figure}
\includegraphics[width=0.23\textwidth]{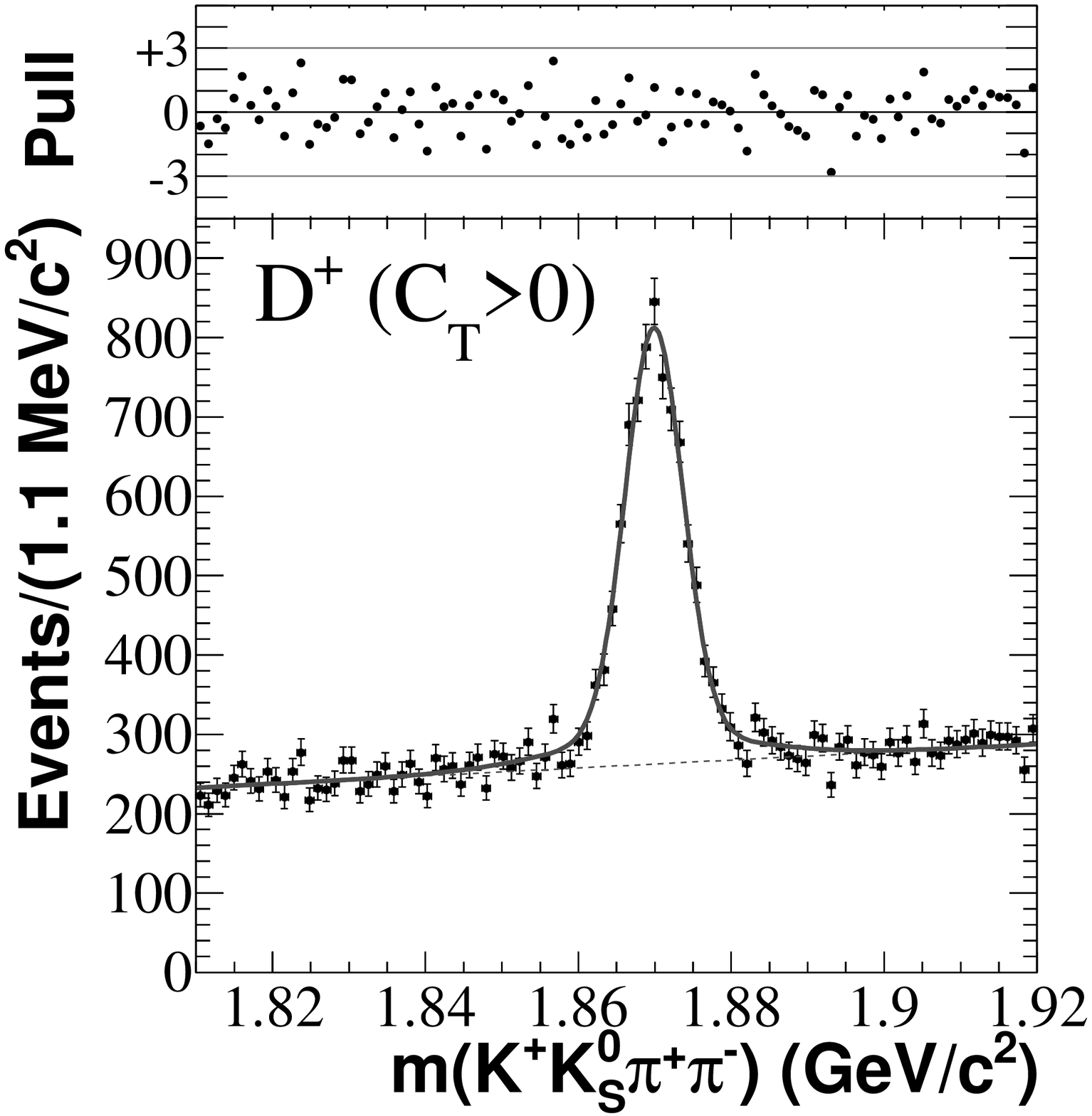}
\includegraphics[width=0.23\textwidth]{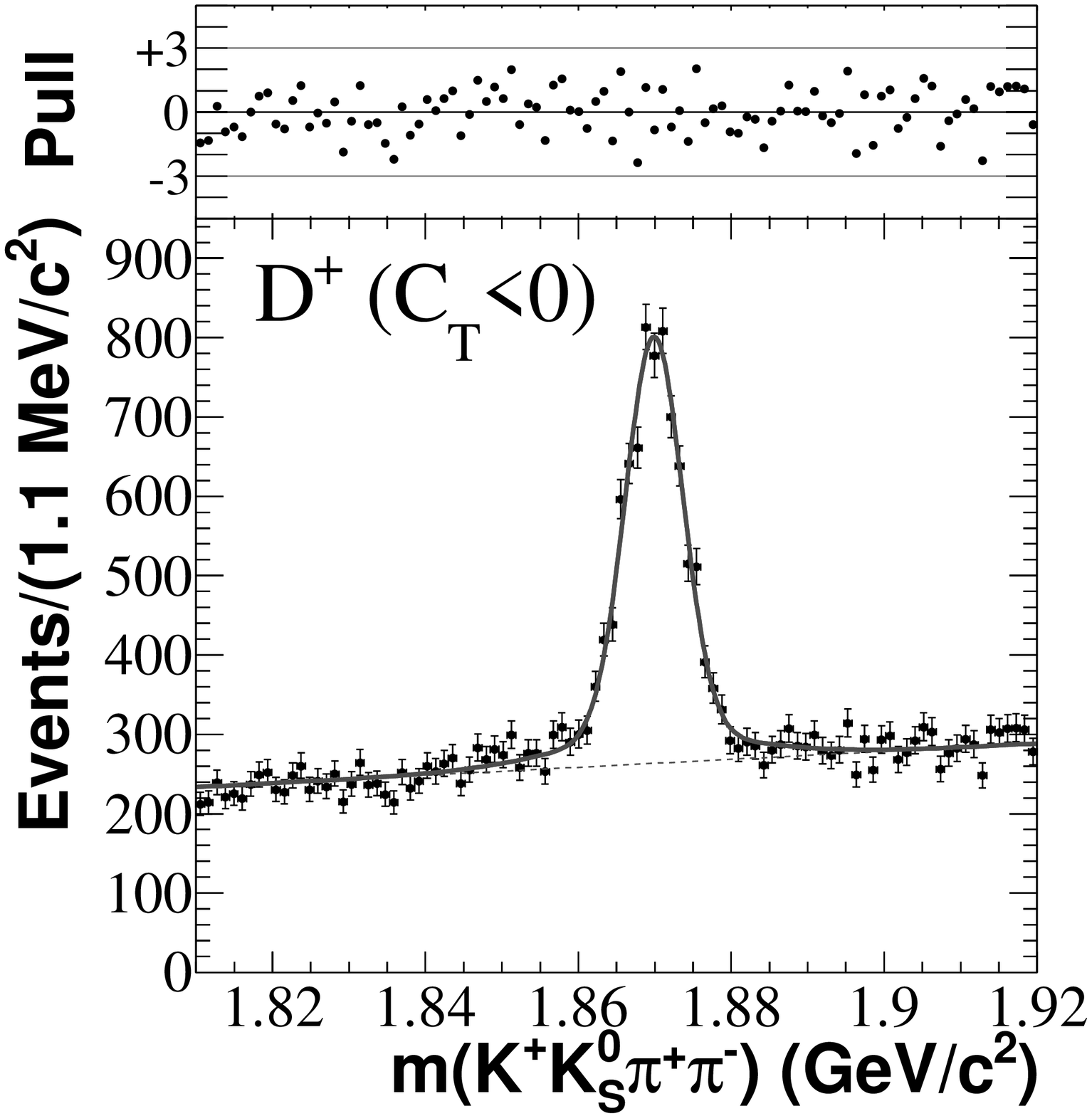}
\includegraphics[width=0.23\textwidth]{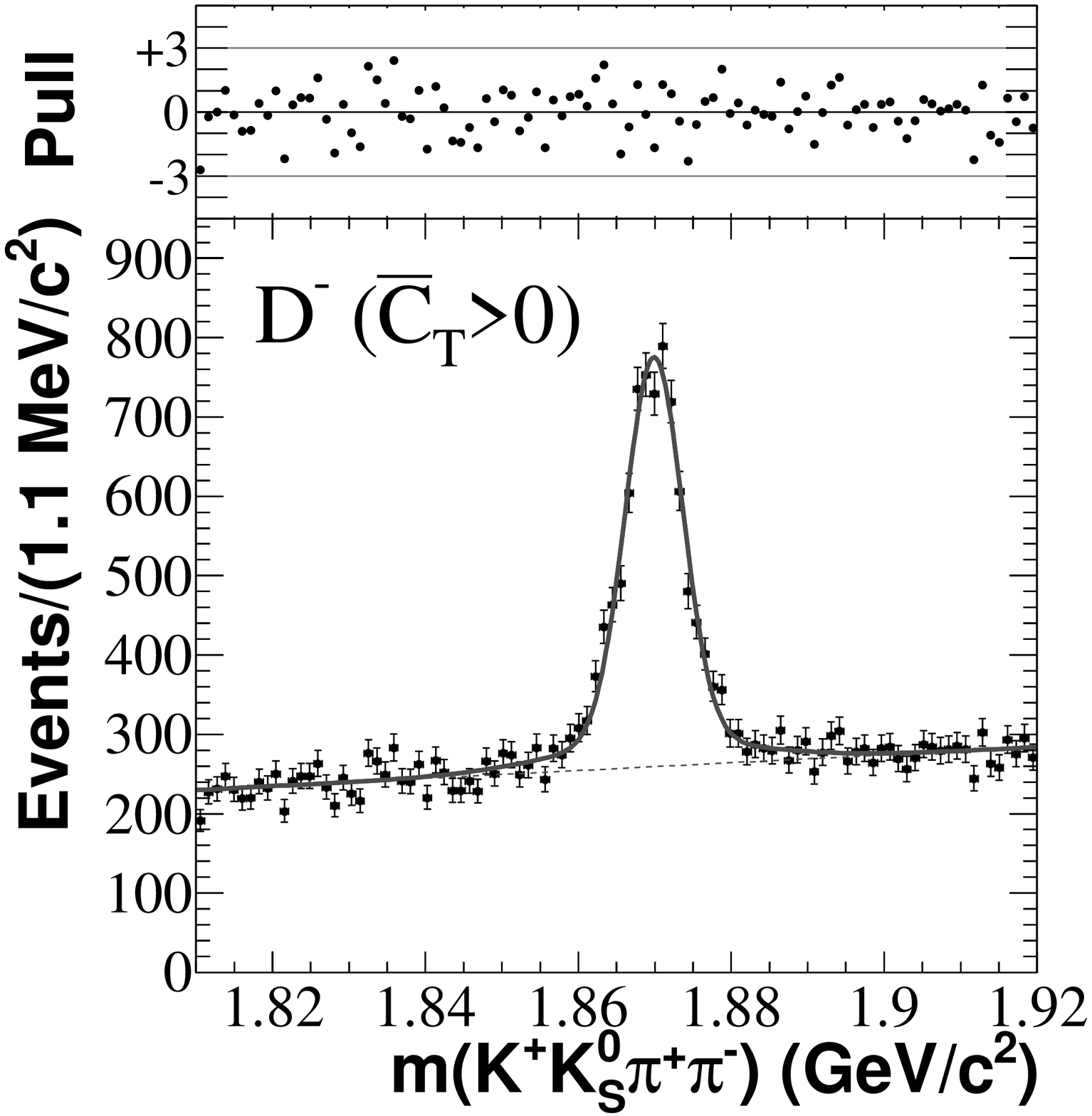}
\includegraphics[width=0.23\textwidth]{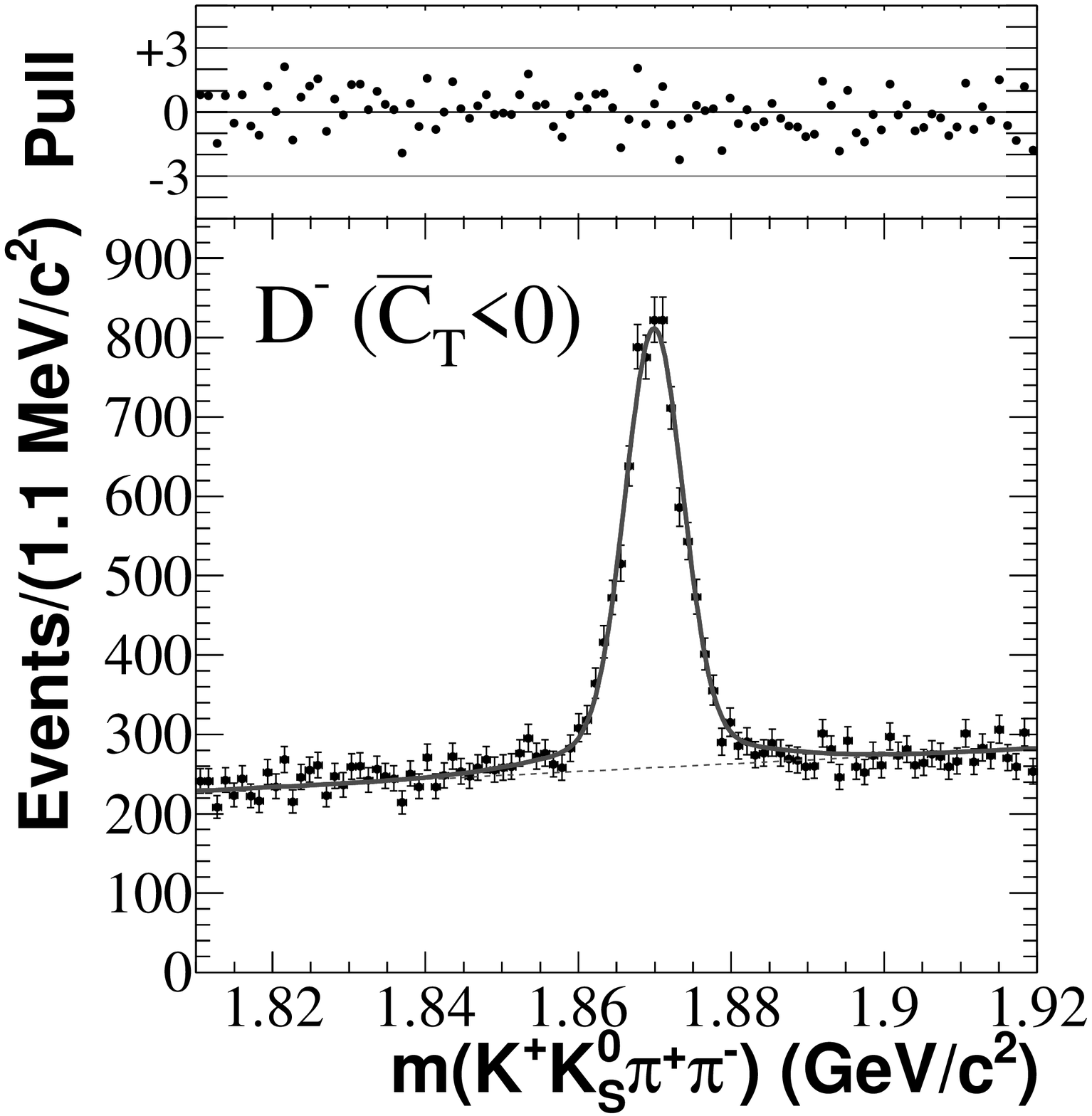}
\caption{\label{fig:DpFit} Fits to the four \DpDecay data subsamples. The pull values are shown above each mass distribution. The $\chi^2/n_{dof}$ values from the fit are 1.07 (\Dp, $\Ct>0$), 1.10 (\Dp, $\Ct<0$), 1.19 (\Dm, $\Ctb>0$), and 0.95 (\Dm, $\Ctb<0$).}
\end{figure}

\begin{figure}
\includegraphics[width=0.23\textwidth]{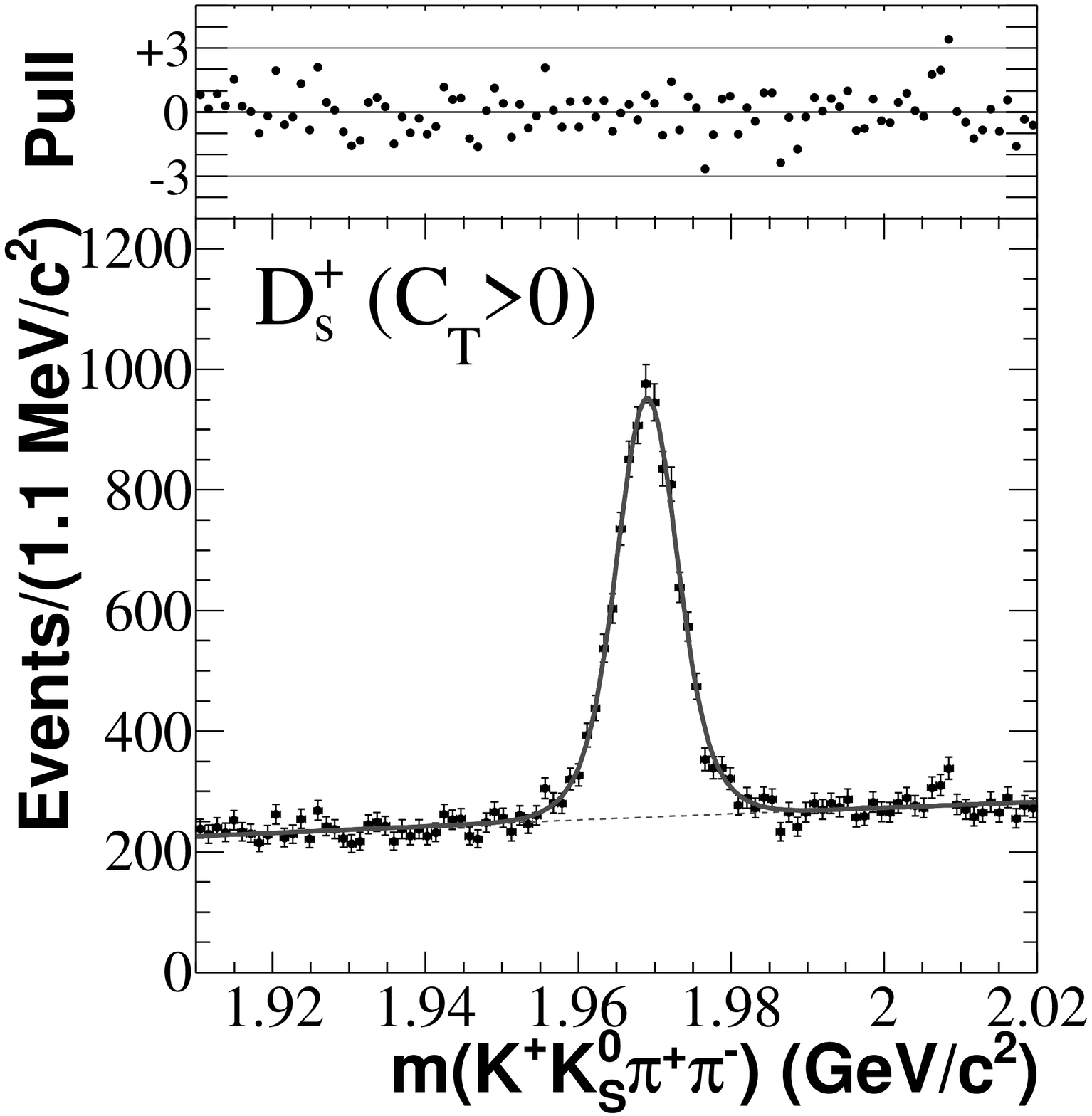}
\includegraphics[width=0.23\textwidth]{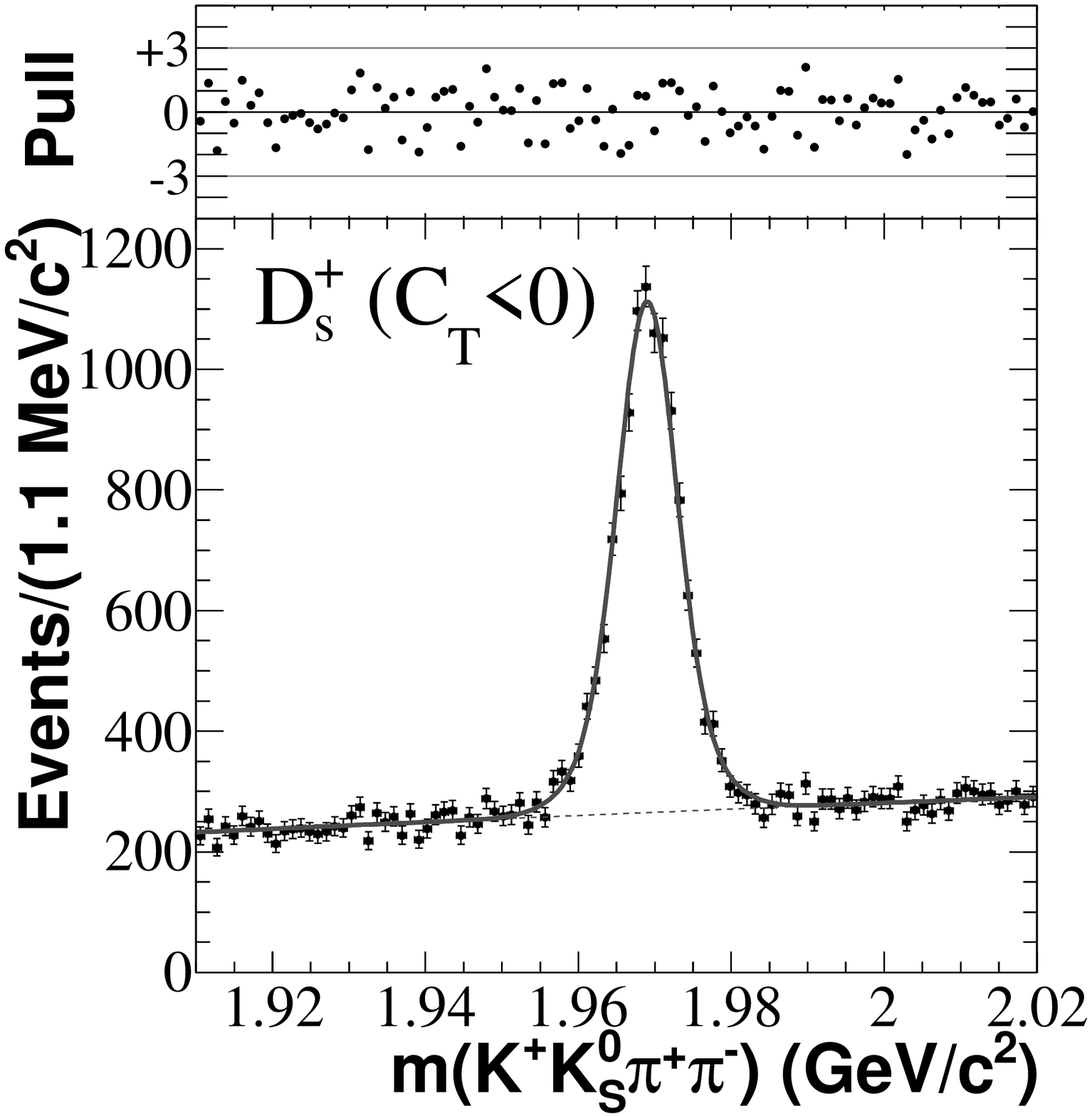}
\includegraphics[width=0.23\textwidth]{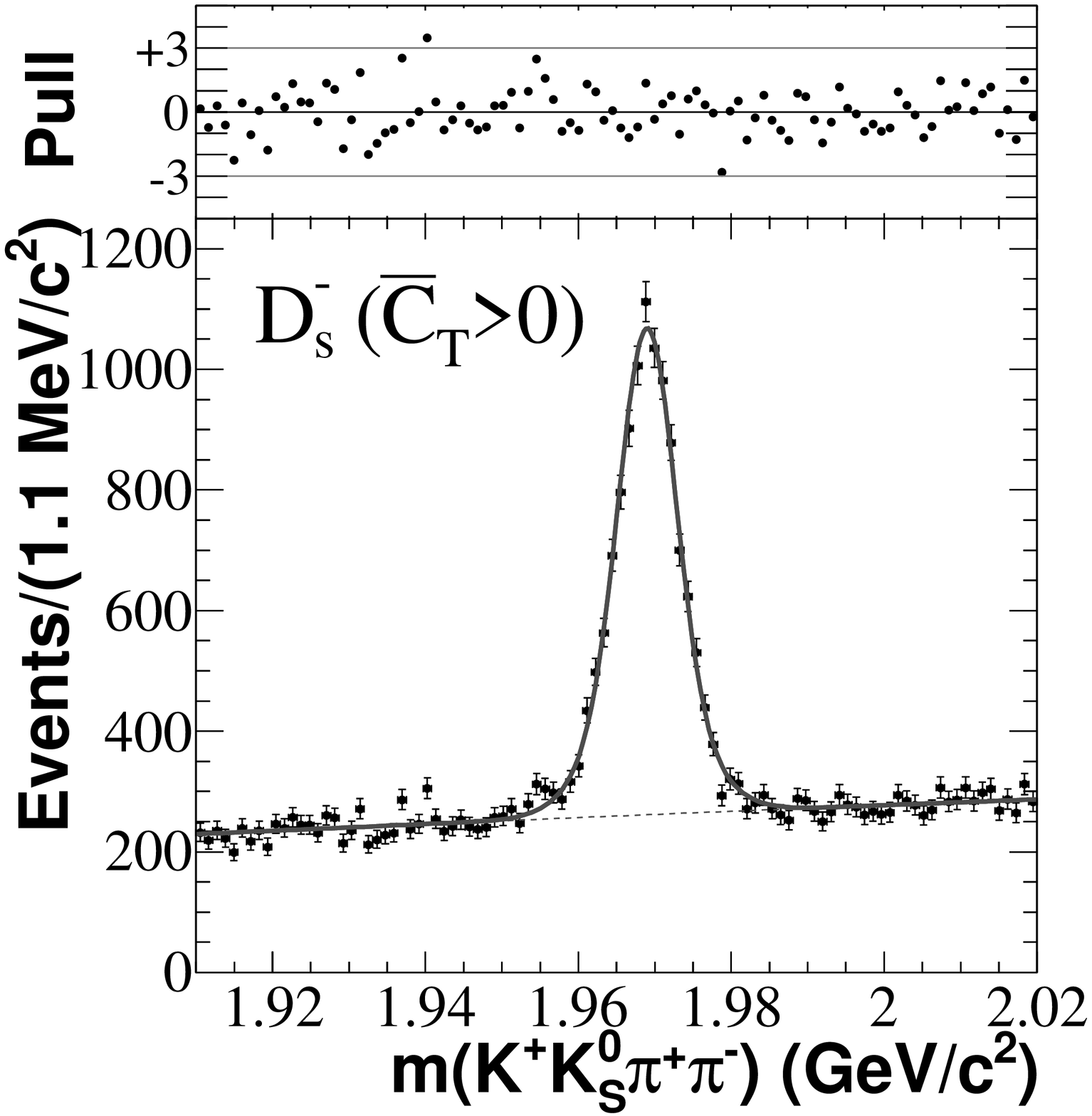}
\includegraphics[width=0.23\textwidth]{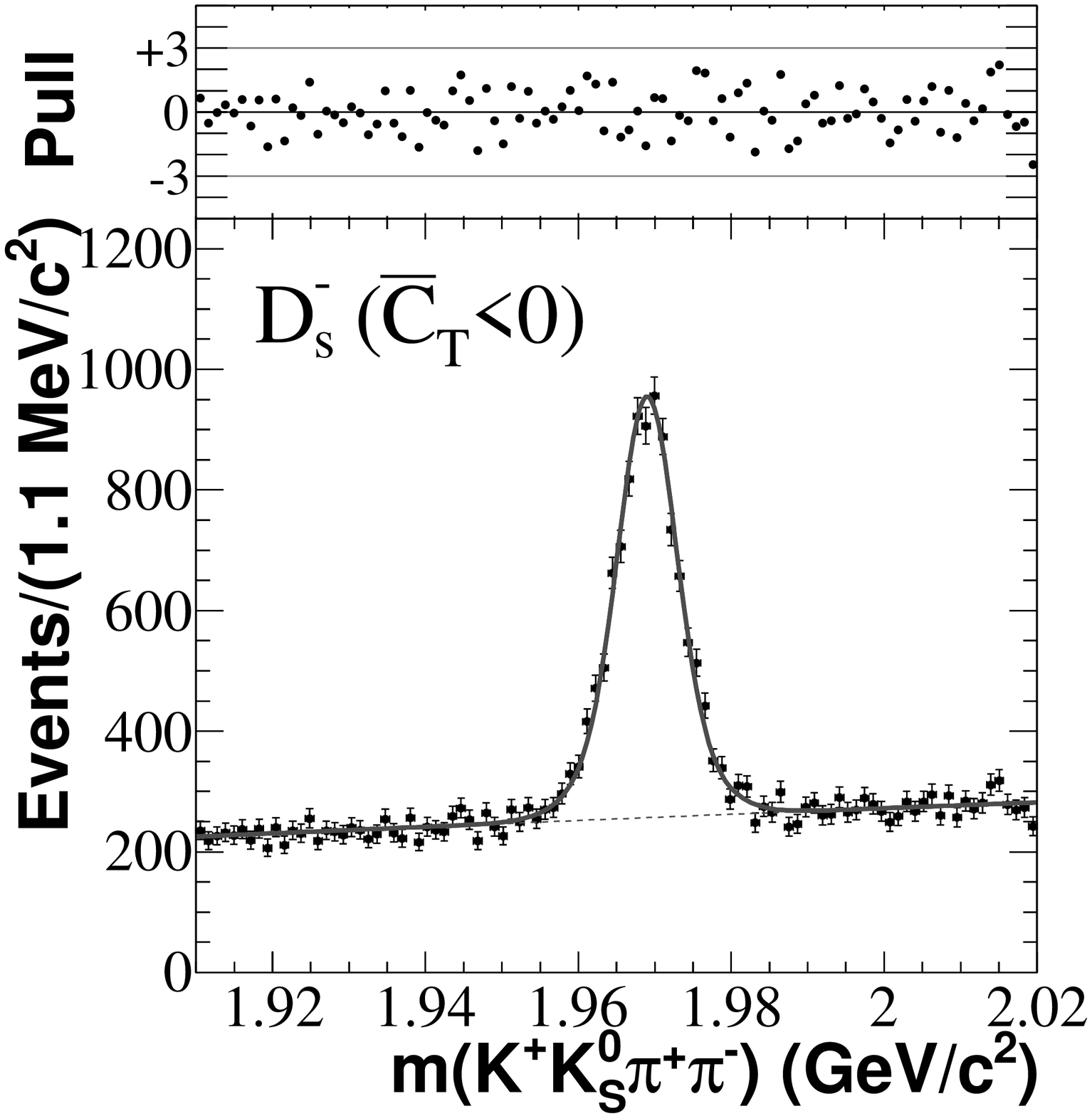}
\caption{\label{fig:DsFit} Fits to the four \DsDecay data subsamples. The pull values are shown above each mass distribution.
The $\chi^2/n_{dof}$ values from the fit are 1.05 (\Ds, $\Ct>0$), 1.03 (\Ds, $\Ct<0$), 1.15 (\Dsm, $\Ctb>0$), and 1.02 (\Dsm, $\Ctb<0$).}
\end{figure}

The fit results for the four data samples are shown in Figs.~\ref{fig:DpFit} and~\ref{fig:DsFit}.
Using Eq.~(\ref{eq:Atv}) we obtain the \T violation parameter values:
\begin{align}
\Atv(\Dp) &= ( -12.0 \pm  10.0_{\stat} \pm 4.6_{\syst} ) \times 10^{-3}
\end{align}
and
\begin{align}
\Atv(\Ds) &= ( -13.6 \pm 7.7_{\stat} \pm 3.4_{\syst} ) \times10^{-3}.
\end{align}
For comparison, the value obtained for \Dz decay was~\cite{delAmoSanchez:2010xj}
\begin{align}
\Atv(\Dz) &= ( +1.0 \pm  5.1_{\stat} \pm 4.4_{\syst} ) \times 10^{-3}.
\end{align}

The sources of systematic uncertainty considered in this analysis are listed in Table~\ref{tab:systDpDs}, and
were derived as follows:
\begin{table*}
\caption{\label{tab:systDpDs} Systematic uncertainty evaluation for \Atv, \At and \Atbar in units of $10^{-3}$ for \DpDecay and \DsDecay.}
\begin{ruledtabular}
\begin{tabular}{lcccccc}
Effect					&\Atv(\Dp)	&\At(\Dp) 	&\Atbar(\Dm) 	&\Atv (\Ds)&\At(\Ds) 	&\Atbar(\Dsm)\\
\hline
(1) Reconstruction			& 2.1		& 2.8		& 1.3 		& 0.7		& 1.0		& 1.3\\
(2) Likelihood ratio 			& 1.1 	& 3.4		& 5.6 		& 2.5 	& 7.8		& 8.2\\
(3) Fit model 				& 1.3 	& 1.1		& 1.5 		& 0.1 	& 0.8		& 0.7\\
(4) Particle identification 		& 3.7 	& 3.3		& 4.1 		& 2.2 	& 2.5		& 6.7\\
\hline
Total						& 4.6		& 5.7		& 7.2			& 3.4		& 8.3		& 10.7\\
\end{tabular}
\end{ruledtabular}
\end{table*}

\begin{enumerate}
\item 
We checked for possible asymmetries resulting from the detector response using large statistics signal MC samples in which the \Dps decays uniformly over phase space.
These events are then weighted according to the resonant structures observed in the data (the resonances that contribute most are $\rho^0\to\pip\pim$, $K^{*0}\to\Kp\pim$, and $K^{*-}\to\KS\pim$). 
Small variations with respect to the generated values are included in the evaluation of the systematic uncertainties.
Using the same samples,
we studied the effect of the forward-backward asymmetry caused by the interference between the electromagnetic current amplitude $\epem\to\gstar\to\ccbar$ and the weak neutral current amplitude $\epem\to\Z\to\ccbar$. 
This interference results in a $D^+_{(s)}/D^-_{(s)}$ production asymmetry that varies linearly with the cosine of the quark production angle $\theta^*$, with respect to the $e^-$ direction. 
Since the \babar\ detector is asymmetric, the final $D^+_{(s)}$ and $D^-_{(s)}$ yields are not equal. 
To include this asymmetry in the MC samples, we weighted them for the $\cos\theta^*$ dependence measured in a previous analysis~\cite{delAmoSanchez:2011xx}.
This study showed that the forward-backward asymmetry does not affect our measurements.
\item We modified the likelihood-ratio selection criteria, and considered the observed deviations from the central parameter values as sources of systematic uncertainty. 
\item In order to check for final state radiation effects, we modified the fitting model by allowing the second Gaussian which describes the signal to have a free mean value.
The background description was also modified by using higher order polynomials.
\item The particle identification algorithms used to identify kaons and pions were modified to more stringent or looser conditions in different combinations.
\end{enumerate}
In the evaluation of the systematic uncertainty for each category, we keep the largest deviation from the reference value, and assume that the uncertainty is symmetric.
It should be noted that the systematic uncertainty on \Atv is not evaluated as the sum in quadrature of the errors on \At and \Atbar. 
Instead, it is evaluated directly from the deviation of \Atv resulting from the fits.
This is why the error from the likelihood ratio or from particle identification is much smaller for \Atv than would be expected from the uncertainties on \At and \Atbar.

In conclusion, we have searched for \CP violation using \todd correlations in high statistics samples of Cabibbo-suppressed \DpDecay and Cabibbo-favored \DsDecay decays.
We obtained $T$-violating asymmetries consistent with zero for both \Dp and \Ds decays with sensitivities of $\approx$ 1.0 \% and $\approx$ 0.8 \%, respectively. We found that possible final-state-interaction effects in the $\Kp\KS\pip\pim$ final state are larger for \Ds decay than for \Dp decay.

We are grateful for the excellent luminosity and machine conditions
provided by our \pep2\ colleagues, 
and for the substantial dedicated effort from
the computing organizations that support \babar.
The collaborating institutions wish to thank 
SLAC for its support and kind hospitality. 
This work is supported by
DOE
and NSF (USA),
NSERC (Canada),
CEA and
CNRS-IN2P3
(France),
BMBF and DFG
(Germany),
INFN (Italy),
FOM (The Netherlands),
NFR (Norway),
MES (Russia),
MICIIN (Spain), and
STFC (United Kingdom). 
Individuals have received support from the
Marie Curie EIF (European Union),
the A.~P.~Sloan Foundation (USA)
and the Binational Science Foundation (USA-Israel).




\begin{thebibliography}{99}
\bibitem{conj}
The use of charge-conjugate reactions is implied throughout.
\end{thebibliography}
\end{document}